\documentclass[12pt]{iopart}
\usepackage{color}
\usepackage{amsfonts}
\usepackage{amssymb}
\usepackage{multirow}
\usepackage{epsfig}

\def\mn{{\mu\nu}}
\def\M{{\cal M}}

\def\b{\beta}

\def\m{\mu}
\def\n{\nu}
\def\r{\rho}
\def\L{\Lambda}
\def\A{{\cal A}}
\def\B{{\cal B}}
\def\S{{\cal S}}
\def\prd{Phys.\ Rev.\ D}
\def\plb{Phys.\ Lett.\ B}

\def\Last{\tilde{\Lambda}_\ast}
\def\Gast{\tilde{G}_\ast}
\def\fast{\tilde f^\ast_2}

\begin{document}
\rightline{PI-QG-185}

\begin{center}

\title{Inflationary solutions in asymptotically safe $f(R)$ theories}

\vspace{1.4cm}
\author{A.~Bonanno$^{1,2}$, A.~Contillo$^{3,4}$, R.~Percacci$^{3,4,5}$}
\vspace{0.5cm}
\address{$^1$INAF - Osservatorio Astrofisico di Catania,
Via S.Sofia 78, I-95123 Catania, Italy\\
$^2$INFN, Sezione di Catania, I-95123 Catania, Italy\\
$^3$SISSA, Via Bonomea 265, I-34136 Trieste, Italy\\
$^4$INFN, Sezione di Trieste, I-34127 Trieste, Italy\\
$^5$Perimeter Institute, 31 Caroline St. N, Waterloo, Ontario N2J 2Y5, Canada}

\end{center}

\vspace*{0.6cm}


\begin{abstract}
We discuss the existence of inflationary solutions 
in a class of renormalization group improved polynomial $f(R)$ theories,
which have been studied recently in the 
context of the asymptotic safety scenario for quantum gravity.
These theories seem to possess a nontrivial ultraviolet fixed point,
where the dimensionful couplings scale according to their
canonical dimensionality.
Assuming that the cutoff is proportional to the Hubble parameter,
we obtain modified Friedmann equations which admit
both power law and exponential solutions.
We establish that for sufficiently high order polynomial
the solutions are reliable, in the sense that considering
still higher order polynomials is very unlikely to change the solution.
\end{abstract}

\maketitle

\section{Introduction}

The idea that a perturbatively nonrenormalizable 
theory could be consistently defined in the UV limit at a nontrivial fixed point (FP),
often called ``asymptotic safety'', is theoretically very attractive, 
especially when applied to quantum gravity \cite{weinberg}.
Much progress in this direction has come from the direct 
application of Renormalization Group (RG) techniques to gravity. 
A particularly useful tool has been the non-perturbative 
Functional Renormalization Group Equation (FRGE) \cite{wetterich},
defining an RG flow on a theory space which consists of all diffeomorphism invariant
functionals of the metric $g_{\mu\nu}$. 
It defines a one parameter family of effective field theories with actions 
$\Gamma_k(g_{\mu\nu})$ depending on a coarse graining scale (or ``cutoff'') $k$,
and interpolating between a ``bare'' action (for $k\to\infty$)
and the ordinary effective action (for $k\to 0$).
When applied to the Einstein-Hilbert action the FRGE yields beta functions
\cite{1998PhRvD..57..971R,1998CQGra..15.3449D} 
which have made possible detailed investigations of the scaling
behavior of Newtons's constant 
\cite{1999PThPh.102..181S,
2002PhRvD..65b5013L,
2002PhRvD..66b5026L,
2002CQGra..19..483L,
2002PhRvD..65f5016R,
2002PhRvD..66l5001R,
2004PhRvL..92t1301L,
2005JHEP...02..035B,
2009GReGr..41..983R,
2009PhRvD..79j5005R,
2009arXiv0904.2510M,
2009arXiv0905.4220M}.
It has been shown quite convincingly that the dimensionful Newton 
constant $G$ is {\em antiscreened} at high energies,
a behavior that eventually leads to the UV FP
that is necessary for asymptotic safety.
These analyses have then been enlarged to include
matter \cite{2003PhRvD..67h1503P} and 
a growing number of purely gravitational operators in the action.
Truncations involving terms quadratic in curvature have been considered
in \cite{2006PhRvL..97v1301C,bms1,bms2,Niedermaier:2009zz}; 
even higher operators could be included restricting
oneself to powers of the Ricci scalar \cite{cpr1,ms1,cpr2}.
It has also been seen in the latter studies that the dimension of the 
basin of attraction of the FP (the so called ``UV critical surface'') 
is likely to be equal to three.
Reviews of this work have appeared in \cite{reviews}.

The RG flow of the effective average action, obtained by different truncations 
of theory space, has been the basis of various investigations of 
``RG improved" black hole \cite{bh,bh2}
and cosmological \cite{cosmo1,cosmofrank,cosmo2,elo,esposito, rw1, rw2} spacetimes. 
In particular, very recently it has been shown that the ``RG improved''
Einstein equations admit (power-law or exponential) inflationary solutions
and that the running of the cosmological constant can account
for the entire entropy of the present universe in the massless sector 
\cite{2007JCAP...08..024B,2008JPhCS.140a2008B} 
(see \cite{claqg08} for an extended review.)

These works were based on the following logic \cite{rw1}.
If we want to study the quantum evolution of the cosmic scale factor,
we should in principle use the full effective action $\Gamma(g_\mn)$, 
which, as we mentioned above,
coincides with $\Gamma_k(g_\mn)$ in the limit $k\to 0$.
However, our knowledge of this functional is rather poor.
One way of gaining some traction on this issue is to observe 
that the Hubble parameter appears as a mass in propagators.
Thus, the contributions of quantum fluctuations with wavelenghts 
greater than $H^{-1}$ are suppressed.
As a result, the functional $\Gamma_k$ at $k$ comparable
to $H$ should be a reasonable approximation for the same functional at $k=0$.
We do not know $\Gamma_k$ much better than the full effective action,
but we can easily calculate the dependence of some terms in $\Gamma_k$ on $k$.
By doing so we effectively take into account nonlocal terms
that would be very hard to calculate otherwise.
This is analogous to what happens in the discussion of the
Coleman-Weinberg potential: there we restrict ourselves to quartic potentials
but we identify the RG scale with the field itself.
Since the quartic coupling runs logarithmically,
this is equivalent to having a term $\phi^4\log(\phi^2)$
in the effective action.

Previous investigations along these lines, in particular \cite{2007JCAP...08..024B},
have been based on the Einstein-Hilbert (EH) truncation.
It is important to establish that the results obtained there 
persist when further operators are included in the truncation.
We know that at the FP the coefficients of these terms are not
very small, but their presence does not seem to affect the values
of the cosmological constant and Newton's constant too much.
In other words, the FP that is is seen in the EH truncation
seems to be robust.
The question then is to see if this stability of the FP against 
the inclusion of new terms 
is reflected in the stability of the corresponding solutions.
This question is important because
the values of the couplings at the FP are fixed and as a consequence
there are no free parameters to be varied.
We will see that the (power-law or exponential) inflationary solutions 
are indeed stable against the inclusion of new terms, but establishing
this fact requires including a rather large number of terms.

A somewhat different perspective on this subject has appeared 
in \cite{weinberginflation}. We will discuss in the conclusions
the relation of that approach to the one adopted here.

Our work is organized as follows. 
First we discuss the equations of motion in the presence of running couplings
(section II), then we specialize the equations to Friedmann Robertson Walker (FRW)
backgrounds (section III).
Section IV is the core of the paper, where we explain in detail our procedure.
The solutions are described in section V.
Section VI defines a different ``RG improvement'' procedure
and discusses the corresponding results.
Section VII is devoted to a discussion of the apparent energy
nonconservation in the ``RG improved'' theory.
In section VIII we compare our approach to other work in the literature
and we discuss the main aspects of this work that need further development.

\section{RG improved gravitational dynamics}

The effective action for the metric containing the effect
of matter and graviton fluctuations is a complicated functional
consisting of all possible integrals of scalar functions constructed with curvatures
and covariant derivatives of curvatures.
If the effective action is evaluated at some reference energy scale $k$,
then the couplings appearing in this effective action will in general
depend upon $k$. The dependence of the couplings on $k$ is described
by $\b$-functions, which collectively describe the RG flow of the theory.
It is practically impossible to say something on the beta functions
of all couplings. In this paper we will consider a subclass of
terms that are somewhat simpler to study than the rest, namely
functions of the curvature scalar $R$. This class of theories has been 
widely studied in the literature, at least at a classical level (see for example \cite{cadefa} for a detailed review).
For the time being, in order to keep it as generic
as possible, we write the action in the form
\begin{equation}
\label{action}
\S=\int d^4x\,\sqrt{|g|}\,F(R)
\end{equation}
where $F$ is some function of the scalar curvature.
Note that even though we will refer to $S$ as ``the action'', 
we always mean ``the effective average action'', meaning that fluctuations
of the fields with momenta greater than $k$ have already been integrated out.
From (\ref{action}) we obtain the equations of motion in the form
\begin{equation}
\label{motiongeneral}
E_{\mu\nu}=\frac{1}{2}T_{\mu\nu}
\end{equation}
where
\begin{eqnarray}
\label{lhs}
E_{\mu\nu}&=&-\frac{1}{\sqrt{|g|}}\frac{\delta S}{\delta g_{\mu\nu}}\nonumber\\
&=&F'(R)R_{\mn}-\frac{1}{2}F(R)g_{\mn}-\nabla_\mu\nabla_\nu F'(R)+g_{\mu\nu}\nabla^\rho\nabla_\rho F'(R)
\end{eqnarray}
and we have written the energy-momentum tensor of matter on the right hand side.
A prime stands for the derivative with respect to $R$.
We want to use these equations to describe the cosmological evolution
of the early universe.

The ``RG improved'' cosmological equations will be obtained by
replacing the gravitational couplings (cosmological constant,
Newton's constant etc.) by running couplings.
Insofar as the cutoff scale is identified with a function of
the metric, and the metric is itself a function of the coordinates, 
the running couplings will also be functions of the coordinates.
We will discuss some other general consequences of this assumption in section VII;
for the time being we just proceed assuming that the function 
$F$ appearing in (\ref{motiongeneral})
is built with such coordinate-dependent couplings, and that this
dependence has to be taken into account when taking its derivatives.
To be more explicit, if we assume that $F(R)$ can be represented by a series
of the form 
\begin{equation}
\label{seriesF}
F(R)=\sum_{i=0}^\infty g_i R^i
\end{equation}
then
$$
\nabla_\mu F'=\sum_i i\left[(i-1)g_iR^{i-2}\nabla_\mu R+\nabla_\mu g_i R^{i-1}\right]
$$
While this seems to us the most natural and correct procedure, it is not
a priori obvious that leaving out the last term (i.e. performing the ``RG improvement''
after having taken the derivatives) would necessarily be wrong.
In order not to burden the reader with a doubling of all results,
in most of this paper we shall follow the former procedure,
and then devote section VI to a discussion of the results using the latter.
Fortunately, we will see that several results are largely independent of this choice.

In the following we will parametrize the action as follows:
\begin{equation}
\label{parametrization}
F=\frac{1}{16\pi G}(f(R)-2\L)
\end{equation}
Then (\ref{motiongeneral}) takes the following form
\begin{eqnarray}
\label{motionext}
&&f'(R)R_{\mn}-\frac{1}{2}f(R)g_{\mn}
-\nabla_\mu\nabla_\nu f'(R)+g_{\mu\nu}\nabla^2 f'(R)
+\frac{\nabla_\mu G}{G}\nabla_\nu f'(R)
\nonumber\\[2mm]
&&+\frac{\nabla_\nu G}{G}\nabla_\mu f'(R)-2\frac{\nabla^\rho G}{G}\nabla_\rho f'(R)g_{\mu\nu}
-f'(R) G\Big ( \nabla_\mu\nabla_\nu \frac{1}{G}-g_{\mu\nu} \nabla^2 \frac{1}{G} \Big)\\[2mm]
&&=8\pi GT_{\mn}-\L g_{\mn}\nonumber
\end{eqnarray}
We will set $f(R)$ to be a polynomial of degree $n$, 
and write it as
\begin{equation}
\label{polyf}
f(R)=\sum^n_{i=1}f_iR^i
\end{equation}
with $f_1=1$ by definition.
Again, it is understood that when derivatives act on $f(R)$,
also the couplings $f_i$ have to be derived.

\section{RG improved Friedmann equations}

As we are interested in the cosmological evolution, we
specialize to a spatially flat Friedmann-Robertson-Walker metric and take
$T^{\m}_{\phantom{\m}\n}=diag(-\r,\,p,\,p,\,p)$ to be the energy momentum
tensor of an ideal fluid with equation of state $p(\r)=w\r$, where $w\neq-1$ is a constant.
Normally at very high energy it is natural to assume $w=1/3$, and we will mostly do so.
However, this is just an approximate description of the matter content of the Universe.
As we shall discuss in section VII, due to the coarse graining the energy
momentum tensor could have unusual properties and the effective
$w$ could be different from its classical value.

In a FRW cosmology with scale factor $a(t)$ we can write both $G_{\mn}$ and
$R_{\mn}$ in terms of the Hubble rate $H(t)=\dot{a}(t)/a(t)$. In particular, we
have
\begin{eqnarray}
\label{various}
&&G_{tt}  =  3H^2, \;\;\;\;\;R_{tt}=-3\,(\dot{H}+H^2),  \;\;
R_{;tt}=\ddot R\nonumber\\[2mm] &&R=R^{\m}_{\phantom{\m}\m}=6(\dot{H}+2H^2),
\;\;\;\; \Box R=-\ddot R- 3H\dot{R}
\end{eqnarray}
so that the $(tt)$-component and (minus) the trace of (\ref{motionext}) become
\begin{eqnarray}
\label{e1}
\,\A(H) & = & 8\pi G\r+\L\\
\label{e2}
\,\B(H) & = & 8\pi G\r\,(1-3w)+4\L
\end{eqnarray}
where
\begin{eqnarray}
\label{Adef}
\A(H)&=&-3(\dot H+H^2) f'+ 3H\dot{f'}+\frac{1}{2}f-3H\frac{\dot G}{G}f'
\\
\nonumber
\B(H)&=&-6(\dot H+2H^2)f'+2f+3\ddot{f'}+\left(9H
-6\frac{\dot G}{G}\right)\dot{f'}\\
\label{Bdef}
&&-3f'\frac{G\ddot G-2\dot G^2+3HG\dot G}{G^2}
\end{eqnarray}

One can eliminate $\rho$ from (\ref{e2}), thus obtaining an equation
that determines $a(t)$, while (\ref{e1}) is used to determine $\rho$:
\begin{eqnarray}
\label{eqn2sqr}
\B(H) &=& (1-3w)\A(H)+3(1+w)\L
\\
\phantom{aaaa} 
\label{eqn2rho}
\rho&=&\frac{1}{8\pi G} (\A(H)-\L)\;.
\end{eqnarray}

The equations for the case without matter can be obtained by setting $\rho=0$ in (\ref{e1},\ref{e2}).
Then $a(t)$ can be obtained by solving
\begin{equation}
\label{nomatter}
4\A(H)=\B(H)
\end{equation}
while either (\ref{e1}) or (\ref{e2}) provide an additional
equation that involves $\Lambda$.
As we shall see, this system is quite constraining.

\section{Cosmology in the fixed point regime}

\subsection{Fixed point action}

The RG flow for $F(R)$ theories of gravity has been studied in
\cite{cpr1,cpr2,ms1}.
One can actually derive a beta functional for the entire function $F$,
but the corresponding FP equation is very complicated, 
so the analysis has been done by expanding $F$ in Taylor series.
It has proved possible to study truncations involving up to
eight powers of $R$.
We now recall the results of this analysis, in the parametrization
provided by equations (\ref{parametrization}) and (\ref{polyf}).

The asymptotic safety scenario posits that in the ultraviolet the couplings 
reach a fixed point.
This statement has the obvious meaning when applied to dimensionless
couplings such as the electromagnetic coupling, or the quartic
coupling in scalar field theory.
In the case of dimensionful couplings, it means that they must
tend to constant values {\it when measured in units of the cutoff}.
We call $k$ the cutoff, and we denote by a tilde the ratio of any quantity by the
cutoff raised to the canonical dimension of that quantity.
Thus quantities with a tilde are by definition dimensionless.
For example $\tilde R=R/k^2$ is the curvature measured in cutoff units, and
\begin{equation}
\label{dimless}
\tilde\Lambda=\Lambda/k^2\ ;\quad
\tilde G=Gk^2\ ;\quad
\tilde f_i=f_i k^{2i-2}
\end{equation}
are the couplings measured in cutoff units.
In the following we will refer to them as ``the dimensionless couplings''.
It is paramount to understand that the existence of a fixed point refers not to the
dimensionful couplings such as $\Lambda$, $G$ etc. but to their
dimensionless counterparts $\tilde\Lambda$, $\tilde G$ etc.
As a result, at a fixed point the dimensionful couplings
do run quite violently, namely they
depend on the cutoff precisely as dictated by naive dimensional analysis.
This is to be contrasted to the low energy regime we are familiar with,
where the opposite happens: the dimensionful couplings do not run, and consequently
the dimensionless ones have a very strong classical running.

We now come to the fixed point of $F(R)$ gravity.
The values of the dimensionless couplings at the fixed point, for various polynomial
truncations of order up to ten, are listed in Table 1.
The first eight lines are taken from \cite{cpr2}.
For reasons that will be discussed in section V.B,
these truncations are actually insufficient for our purposes
and we have found it necessary to examine also the truncations $n=9,10$.
All the other details of the calculation (cutoff, gauge etc.)
are the same as in \cite{cpr2}.
We have found only the two fixed points listed in the last two lines
in the table.
We have not calculated the critical exponents that pertain to
these fixed points, so our understanding of their nature is less
complete than for the truncations $n\leq 8$, but the close resemblance
of the values of most couplings is a strong hint that these
are indeed the correct prolongations of the $n=8$ fixed point
to higher truncations
\footnote{One may be worried by the sign flip of $\tilde f^\ast_6$.
In this connection we note that the coefficients arise from
the sum of a large number of terms and that there is nothing to guarantee
their sign. The actual difference
between $\tilde f^\ast_6$ in the truncations $n=8$ and $n=9$ is 0.08; 
similar shifts occur also elsewhere in the table.
}.

\begin{table}
\begin{center}
\begin{tabular}{|c|l|l|r|r|r|r|r|r|r|r|r|}
\hline
 $n$ & 
 $\tilde\Lambda_*$&
 $\tilde G_*$ & 
 \multispan9 \hfil $10^3\times$ \hfil\vline
\\
\hline
 & & & $\tilde f^\ast_2$ & $\tilde f^\ast_3$
& $\tilde f^\ast_4$ & $\tilde f^\ast_5$ & $\tilde f^\ast_6$& $\tilde f^\ast_7$& $\tilde f^\ast_8$
& $\tilde f^\ast_9$& $\tilde f^\ast_{10}$
\\
\hline
1&0.1297 &0.9878 & & & & & & & & &\\
2&0.1294 &1.5633 & -119.0& & & & & & & &\\
3&0.1323 &1.0152 & -35.82& 494.1& & & & & & &\\
4&0.1229 &0.9664 & -13.14& 532.8& 420.0& & & & & &\\
5&0.1235 &0.9686 & -13.11& 471.7& 391.1& 163.0& & & & &\\
6&0.1216 &0.9583 & -6.81& 491.2& 460.9& 172.9& -118.5& & & &\\
7&0.1202 &0.9488 &  1.61& 466.6& 501.8& 288.4& -163.1& -218.6& & & \\
8&0.1221 &0.9589 & -4.225& 413.6& 430.2& 328.1& -56.17& -298.6&-226.3& & \\
9&0.1242 &0.9715 & -13.17& 398.5& 362.9& 297.7& 28.77& -223.2&-219.3& -127.7& \\
10&0.1242 &0.9718 & -13.16& 391.1& 360.7& 307.8& 3.874& -230.1&-229.3& -123.9& 40.32\\
\hline
\end{tabular}
\caption{Position of the FP as a function of $n$, the order of the truncation.
To avoid writing too many decimals, the values of $\tilde f^\ast_i$ have
been multiplied by 1000.}
\end{center}
\label{tab:mytable1}
\end{table}

In this paper we want to use this information in cosmology.
Since the universe becomes hotter and more strongly curved as one proceeds backwards towards
the big bang, it is clear that in order to explain the
dynamics of the early phases of its evolution, we need to have
a theoretical model for the behavior of interactions at high energies.
Asymptotically safe gravity is a model for what happens near the
Planck scale: it posits that all interactions reach the fixed point regime.
So, assuming that gravity is asymptotically safe,
in this paper we want to study the evolution of the cosmic scale factor
in the very early universe, when the (dimensionless) couplings are so close
to their fixed point value that we can actually use
the values give in table I.

\subsection{Cutoff choice}

The RG equations give us the dependence of the couplings on some cutoff
scale $k$, but they do not tell us what $k$ is.
In order to apply them to cosmology, or any other problem, 
one has to identify a physical quantity that acts like a cutoff.
From a Wilsonian point of view, the cutoff is a typical 
energy or momentum scale of the phenomenon under study, 
and one integrates out all fluctuations
of the fields with momenta higher than the cutoff.
But this still falls short of identifying it uniquely.
The best choice of $k$ has to be made
on a case by case basis, analyzing the physical problem at hand.

In our case, as also pointed out in  \cite{weinberginflation}, the Hubble constant $H$ provides a natural infrared cutoff in the loop diagrams of the gravitons (see also \cite{claqg08} for an extended discussion of this point). 
We will therefore assume that the cutoff is
\begin{equation}
\label{cutoff}
k=\xi H
\end{equation}
where $\xi$ is a positive number of order unity 
\footnote{Different choices of cutoff have been considered in the past,
for example $k\sim 1/a(t)$ \cite{florean,bauer} and $k\sim 1/t$
\cite{cosmo1,rw1,rw2}. The choice made here has been considered previously in
\cite{2007JCAP...08..024B}. See also \cite{bab} and \cite{sha}
and more recently \cite{weinstein,brandenberger}.
Note that for a power-law dependence of the scale factor 
the choices $k\sim 1/t$ and $k\sim H$ are equivalent.
}.

The logic that we will use is then as follows:
in the fixed point regime we can replace $k$ by $\xi H$ in (\ref{dimless}).
This turns the dimensionful couplings $\Lambda$, $G$ etc into functions of time.
We then use these expressions in the dynamical equations (\ref{e1}) and (\ref{e2}).
The dimensionful couplings are replaced by powers of $H$, dimensionless
couplings which we take from table I and the free parameter $\xi$.
The occurrence of the time dependent function $H$, where previously there appeared
a constant, obviously complicates the equations significantly.
The resulting equations should give a reasonably good approximate description
of the dynamics of the universe in the fixed point regime.
We then look for de Sitter type solutions
\begin{equation}
\label{exponential}
a(t)=a_0 e^{Ht}\ ;\qquad
H=\mathrm{constant}\ ,
\end{equation}
or power law solutions
\begin{equation}
\label{powerlaw}
a(t)=a_0 t^p\ ;\qquad
H=\frac{p}{t}\ .
\end{equation}

\begin{figure}
\label{fig1}
\includegraphics[width=10cm,height=6cm]{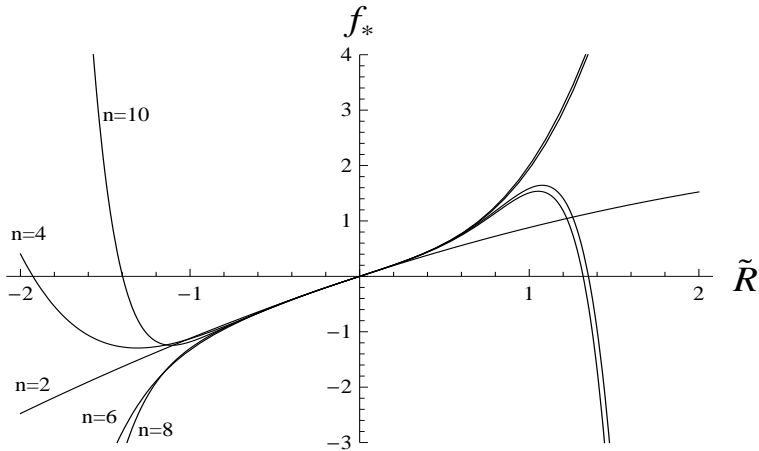}
\caption{Taylor expansion of $\tilde f_*$ around zero for $n=2,4,6,8,10$.
}
\end{figure}

\subsection{Reliability analysis}

One important advantage of having the results of increasingly more complete truncations
is that this gives us some quantitative handle on the reliability of the calculations.
To understand this point let us write 
$$
f(R)=k^2\tilde f(\tilde R)\ ; \qquad
\tilde f(\tilde R)=\sum^n_{i=1}\tilde f_i\tilde R^i
$$
The fixed point polynomial $\tilde f_*(\tilde R)$, 
whose coefficients are given in table I, is plotted in figure 1 for $n=2,4,6,8,10$.
Clearly the approximation becomes more accurate as $n$ increases.
It is worth emphasizing that table I does not give the Taylor expansion
of a fixed function: if that was the case, then a single row of coefficients
would have been enough.
Instead, for each $n$, the truncated fixed point equations give a whole new set of
coefficients, which provide a polynomial approximation for the ``true'' fixed point
function $\tilde f_*$. The fact that the numbers in the columns of table I
do not change too wildly is an encouraging sign that the data in the table do
indeed resemble a Taylor expansion of some function.
As expected, one sees that the shape of the function near the origin is
rather unchanging, and that the uncertainty moves progressively to larger $\tilde R$.
In order to make this quantitatively more precise we shall use the
following method. 
Let $\tilde f^\ast_n$ be the fixed point curve in the truncation $n$.
We say that $\tilde f^\ast_n$ is reliable 
as long as it differs from $\tilde f^\ast_{n+1}$ by less than $\Delta f^*$.
Here we will take $\Delta f^*=0.025$: 
since $\tilde f^*\approx 1$ for $\tilde R\approx 1$,
this seems a reasonable criterion for two successive approximations
being ``near'' to each other in some domain.
Then we find that the truncations are reliable for
\begin{equation}
\label{silvia}
\tilde R\lesssim c
\end{equation}
where $c$ is given by the following table:
\begin{center}
\begin{tabular}{|c|r|r|r|r|r|r|r|r|r|r|}
\hline
$n$ & 1 & 2 & 3& 4& 5& 6& 7& 8& 9& 10\\
\hline
$c$ & 0.45 & 0.32 & 0.45& 0.87& 1.03& 0.86& 0.76& 0.87& 0.99& 0.99\\
\hline
\end{tabular}
\end{center}
This method cannot give a reliability value for the
highest truncation, so we conservatively assume that 
the $n=10$ solution has the same range as the $n=9$ one.

Using (\ref{various}) and (\ref{cutoff}), equation (\ref{silvia}) implies that
\begin{equation}
\label{bound}
c\xi^2 \gtrsim 12+6\frac{\dot H}{H^2}
\end{equation}
Later on, when we discuss solutions, we will make sure that
they occur within these bounds.

One may wonder how things change when one uses different definitions for the ``nearness'' of two functions.
Of course if one chooses $\Delta f^*$ to be too small, then also the best truncations
become unreliable and conversely if $\Delta f^*$ is too large then all truncations
are reliable. Values of $\Delta f^*$ between 0.01 and 0.1 give similar results.
The values of $c$ in the table have a clear tendency to increase, but they have
also some randomness. This can be decreased by choosing a different criterion for
what one means by nearness, for example defining $c$ by the requirement that the
integral of the modulus of the difference of two successive approximations be
less than a preset value. We think that the simple criterion used above is sufficient
for our purposes.

\section{Cosmological solutions}
\subsection{Einstein-Hilbert truncation}
In order to make contact with \cite{2007JCAP...08..024B}
we discuss first the case $n=1$, corresponding to the Hilbert action
$f(R)=R$.
In this case the expressions (\ref{Adef}) and (\ref{Bdef})
reduce to
\begin{eqnarray}
\A(H)&=&3H^2-3H\frac{\dot G}{G}
\\
\B(H)&=&6(\dot H+2H^2)
-3\frac{G\ddot G-2\dot G^2+3HG\dot G}{G^2}
\end{eqnarray}

It is not difficult to show that power law solutions of the type (\ref{powerlaw})
are obtained for  
\begin{equation}
\label{solp1more}
p=3\,\frac{1\pm\sqrt{1-\frac{2}{3}\,\frac{3-\Last\xi^2}{1+w}}}{3-\Last\xi^2}
\end{equation}
In particular we see that $p$ is real and positive for $\xi$ greater than a critical value.

In reference \cite{2007JCAP...08..024B} a different RG improvement was used:
the running couplings were inserted in Einstein's equations,
in such a way that the $\dot G$ terms were absent.
We defer a detailed discussion of the difference between these procedures to section VI.
For the time being we merely notice that using the approach described 
in \cite{2007JCAP...08..024B}, for $w=1/3$ one obtains 
\begin{equation}
\label{solp1less}
p=\frac{3}{6-2\tilde\Lambda_*\xi^2}
\end{equation}

The exponents are plotted in Fig.2 as functions of $\xi$. 
Note that the positive branch of (\ref{solp1more}) is qualitatively similar
to (\ref{solp1less}) (the short-dashed curve).
Unfortunately both curves are in the region where the truncation
is not reliable according to the criterion (\ref{silvia})
(they are in the grey area).
It is interesting to notice that for $\xi > 4$ 
the negative branch of (\ref{solp1more})
is within the domain of validity of our approximation,
but then the exponent is too small for inflation.

%
\begin{figure}
\label{zano}
\includegraphics[width=10cm,height=6cm]{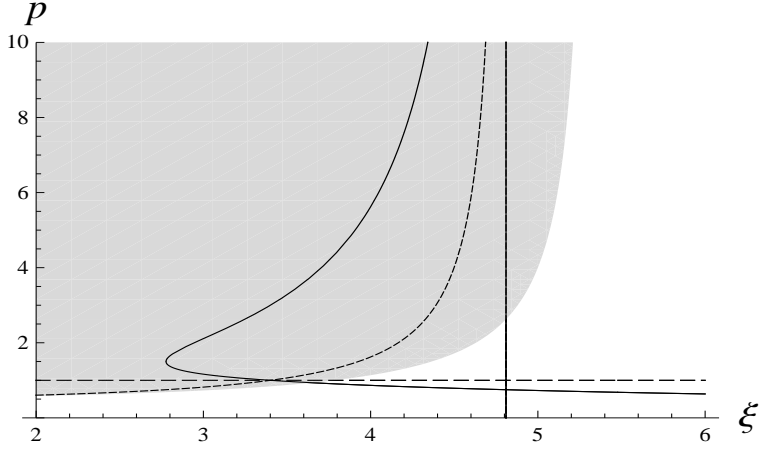}
\caption{Black solid curve: Exponents $p$ of power law solutions as a function of $\xi$
for $n=1$, $w=1/3$; see equation (\ref{solp1more}). 
The short-dashed curve is the solution (\ref{solp1less}), 
found in \cite{2007JCAP...08..024B} and further discussed in section VI.
The long-dashed curve indicates $p=1$. The grey region represents the region excluded by the reliability criterion,
see (\ref{bound}) with $c=0.42$.
}
\end{figure}

\subsection{The case $n=2$}

As another explicit example let us give the form of the equations
when $f=R+f_2 R^2$.
Then we have
\begin{eqnarray}
\A(H)&=&3H^2+18 f_2(2H\ddot H-\dot H^2+6H^2\dot H) 
\nonumber\\
&&
+36\dot f_2H(\dot H+2H^2)
-3H\frac{\dot G}{G}\left[1+12f_2(\dot H+2H^2)\right]
\\
\B(H)&=&6(\dot H+2H^2)+36f_2(H^{(3)}+7H\ddot H+12H\dot H+4\dot H^2)
\nonumber\\
&&
+36\dot f_2(2\ddot H+11H\dot H+6H^3)
+36\ddot f_2(\dot H+2H^2)
\nonumber\\
&&
-72\frac{\dot G}{G}\left[\dot f_2(\dot H+2H^2)+f_2(\ddot H+4H\dot H)\right]
\nonumber\\
&&
-3\frac{G\ddot G-2\dot G^2+3HG\dot G}{G^2}\left[1+12f_2(\dot H+2H^2)\right]
\end{eqnarray}

The power-law exponents $p$ are now obtained from the solution of a cubic equation. Their explicit form reads
\begin{eqnarray}
\label{sol2more}
p_1&=&\frac{36(1+w)\left(18\fast+\xi^2\right)+\frac{2^{4/3}\M_1}{\M_2^{1/3}}+2^{2/3}\M_2^{1/3}}{18(1+w)\xi^2\left(3-\Last\xi^2\right)}\\ \nonumber
p_{2,3}&=&\frac{72(1+w)\left(18\fast+\xi^2\right)-\frac{2^{4/3}(1\pm i\,\sqrt{3})\M_1}{\M_2^{1/3}}-2^{2/3}(1\mp i\,\sqrt{3})\M_2^{1/3}}{18(1+w)\xi^2\left(3-\Last\xi^2\right)}
\end{eqnarray}
where
\begin{eqnarray}
\M_1 &=& 162(1+w)\left(2(1+w)\left(18\fast+\xi^2\right)^2\right.\nonumber\\
&&\left.-\xi^2\left(3(11+3w)\fast+\xi^2\right)\left(3-\Last\xi^2\right)\right)
\end{eqnarray}
and
\begin{eqnarray}\nonumber
 \M_2&=&11664(1+w)^3\left(18\fast+\xi^2\right)^3-8748(1+w)^2\xi^2\left(18\fast+\xi^2\right)\\ \nonumber
&&\times\left(3(11+3w)\fast+\xi^2\right)\left(3-\Last\xi^2\right)\\
&&+52488(1+w)^2\fast\xi^4\left(3-\Last\xi^2\right)^2\nonumber\\
&&+\left[-4\M_1^3+8503856(1+w)^4\left(4(1+w)\left(18\fast+\xi^2\right)^3\right.\right.\\
&&-3\xi^2\left(18\fast+\xi^2\right)\left(3(11+3w)\fast+\xi^2\right)\left(3-\Last\xi^2\right)\nonumber\\
&&\left.\left.+18\fast\xi^4\left(3-\Last\xi^2\right)^2\right)^2\right]^{1/2}\nonumber
\end{eqnarray}

The exponents as functions of $\xi$ are then plotted in figure 3.
It turns out that $p_1$ is negative and therefore uninteresting for cosmology;
The other two solutions are complex in general,
but for $\xi$ greater than some critical value they are real.
We see that these solutions are very similar to the ones found in EH truncation.
Of the two branches, only the lower one is reliable for large $\xi$,
but again it is too small to be of interest for inflation.
The upper branch is not in the region where the truncation is reliable.

\begin{figure}
\includegraphics[width=10cm,height=6cm]{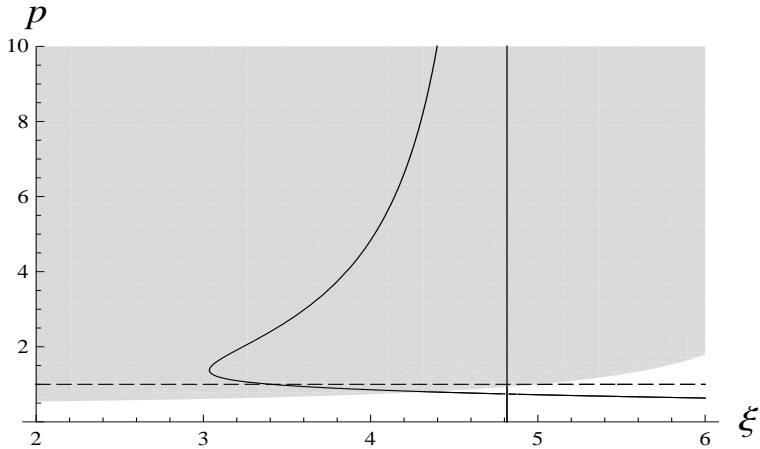}
\caption{Black solid curve: Exponents $p$ of power law solutions as a function of $\xi$
for $n=2$, $w=1/3$; see equation (\ref{sol2more}). 
The dashed curve indicates $p=1$. The grey region represents the region excluded by the reliability criterion,
see (\ref{bound}) with $c=0.24$.
}
\end{figure}

\subsection{The general case: de Sitter solutions}

If we make the ansatz (\ref{exponential}), the terms containing $\dot{f'}$ and $\dot G$ 
in equations (\ref{Adef},\ref{Bdef}) vanish, and one finds 
$$
\B=4\A\ .
$$
Inserting in equations (\ref{e1},\ref{e2}) there follows that $w=-1$, 
i.e. matter must have the same equation of state as the cosmological constant.
We may then as well absorb such matter in the definition of $\Lambda$ and set $\rho=0$.
Thus without loss of generality we will study the exponential solution only in the absence of matter.

Since $R=12H$ is constant, equation (\ref{eqn2sqr}) reduces to 
\begin{equation}
\label{expeq}
Rf'-2f+4\Lambda=0\ .
\end{equation}
This equation had been studied earlier in \cite{barrow}.
We render the equation dimensionless by going to tilde variables
and use the values of the dimensionless couplings given in table I.
Then we solve for $\tilde R$.
Numerically we find solutions for $n=2,3,6,7,8,9,10$.
These are given in the second column of table II.
Note that for any solution $\tilde R_*$, the value of the 
parameter $\xi$ is fixed.
This is because in de Sitter space $R=12 H^2$,
so $R_*=\tilde R_* k^2=\tilde R_* H^2\xi^2$
implies that $\xi^2=12/\tilde R_*$.
The solutions mentioned above would correspond to
the values of $\xi$ shown in the third column in table II.

No real solutions are found in the truncations $n=4,5$.
In order to understand the reason for this consider the coefficients
in table I and observe figure 4 where we have plotted
the left hand side of the equation.
The behavior of the function $\tilde R\tilde f'-2\tilde f$ for large $\tilde R$
is determined by the sign of the coefficient $(n-2)f^\ast_n$.
For $n=2$, this leading term cancels and the equation is 
determined by the linear term which is negative, so that there is a zero.
For $n>2$ the sign of this coefficient is the same
as the sign of $f^\ast_n$.
For $n=3$, it is positive and the curve is bent
upwards but not enough to eliminate the solution.
For $n=4,5$, $f^\ast_4$ and $f^\ast_5$ are
again positive and the curve is bent upwards enough that the
solution disappears.
If this was as far as one could get with the truncation,
then one might conclude that the solution that is present
for $n=2,3$ is a truncation artifact.
In fact the solution occurs at values of $\tilde R$
that are outside the reliable range defined by (\ref{silvia}).
Now, when we add the terms of order 6, 7 and 8, the coefficients
of the highest terms are negative and therefore a solution reappears,
first at a rather large (and hence unreliable) value of $\tilde R$
but then at reasonably small $\tilde R$.
Is this sufficient evidence for the existence of the solution?
If we were to rely only on the results of \cite{cpr2},
where $n=8$ was the highest truncation considered, 
one would feel that the evidence is somewhat inconclusive.
In fact, looking at figure 4, where $n=8$ is represented by the
dashed curve, one can easily imagine that if $f^\ast_9$ was sufficiently 
positive the solution could disappear again.
It is for this reason that we have looked at the fixed point condition
in the cases $n=9,10$.
Luckily $f^\ast_9$ is again negative, and $f^\ast_{10}$ is positive but not very large,
so that a solution exists in all these cases and actually occurs 
within the domain of reliability of the truncations.
Furthermore, the position of the solution seems to be quite stable
for $n=8,9,10$, which suggests that the higher order terms will 
not affect it too much.
The hard lesson that one learns from this is that it may be necessary to go to
very high truncations before one obtains reliable results.

\begin{table}
\begin{center}
\begin{tabular}{|c|l|l|}
\hline
$n$ & $\tilde R_*$ & $\xi$ \\
\hline
1& 0.51897 & 4.8086 \\
2& 0.51743 & 4.8157\\
3& 0.69486 & 4.1557\\
6& 2.14818 & 2.3635\\
7& 0.93384 & 3.5847\\

8& 0.78561 & 3.9083\\
9& 0.75799 & 3.9788\\
10& 0.76922 & 3.9497\\
\hline
\end{tabular}
\caption{Table 2. De Sitter solutions for various truncations. 
When solutions exist, only the smaller one is displayed; no solutions are found for $n=4,5$.}
\end{center}
\label{tab:mytable2}
\end{table}

\begin{figure}
\label{fig1a}
\includegraphics[width=10cm,height=6cm]{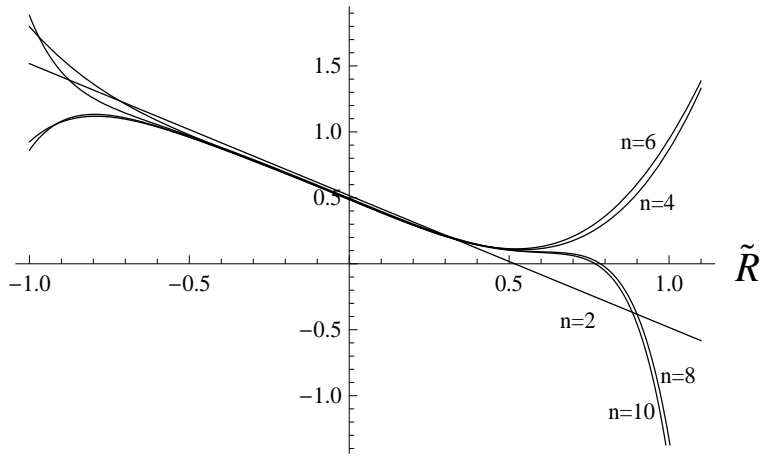}
\caption{The left hand side of equation (\ref{expeq}) as a function of $\tilde R$.}
\end{figure}

\subsection{The general case: power law solutions}

Now we look for power law solutions (\ref{powerlaw}).
In this case the condition (\ref{bound}) implies
\begin{equation}
\label{boundp}
c\xi^2 \gtrsim 12-\frac{6}{p}
\end{equation}
Consider first the case without matter.  Then we have to solve the equations 
\begin{equation}
\A(H)  =  \L \;\;\;\;\;\;3\,\B(H)  =  4\L
\end{equation}
When one uses (\ref{Adef}), (\ref{Bdef}), (\ref{polyf}), (\ref{various}) and (\ref{cutoff}), 
for dimensional reasons these equations reduce to the product of some power of $t$ times a function of $p$ and $\xi$.
Solving them for all $t$ means that $p$ and $\xi$ must satisfy two algebraic equations,
and one expects to find at most isolated solutions.
At least for $n=2$ no solutions were found.

\begin{figure}
\includegraphics[width=10cm,height=6cm]{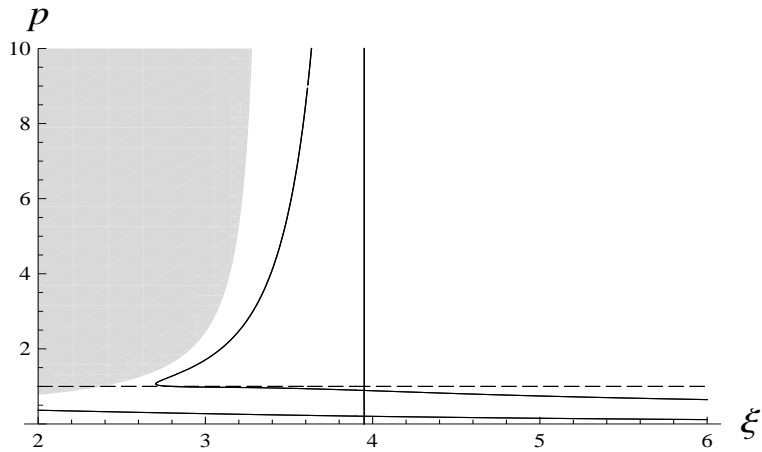}
\caption{Solid curve: numerical solutions for the exponents 
of power law solutions as a function of $\xi$ for $n=10$, $w=1/3$. 
The dashed curve indicates $p=1$. The grey region represents the region excluded by the reliability criterion,
see (\ref{bound}) with $c=1.06$.
}
\end{figure}
Let us now search for power law solutions in the presence of matter.
Things now look better because matter gives us a new degree of freedom
and allows us to find solutions for continuous ranges of values of $\xi$.
Namely, for fixed $\xi$ one can use equation (\ref{eqn2sqr})
to determine the exponent $p$ and then use (\ref{eqn2rho}) to fix $\rho(t)$.
In this way one avoids having to fix $\xi$.
The case $n=2$ has already been discussed in section 5.2.
The cases with $n>2$ cannot be solved analytically,
but solutions exist and can be found numerically.
As before we use $w=1/3$.

Doing this highlights once again the need to go beyond the truncation $n=8$.
For suppose that we only knew the results up to $n=8$.
Then according to our criteria we would have to take
for this truncation the same reliability range as the
$n=7$ truncation, which has a smaller value $c=0.79$.
With this criterion, the whole $n=8$ solution that asymptotes
to de Sitter would be unreliable.
It is only by going to higher $n$ that we can validate
the solution in an acceptable range.
We note that part of these curves lie still to the left of the
reliability limit (\ref{boundp}), but that is the part
that is less interesting physically.
As a confirmation of the results, we show in figure 5
the exponents $p$ in the truncation $n=10$.

\section{A more restrictive RG improvement}

Let us return to equation (\ref{motiongeneral}).
As we mentioned, there is a possible ambiguity concerning
the stage at which one should replace the
usual ``constant'' couplings by time dependent ``running'' couplings.
To illustrate this point, let us consider
for a moment the special case of the Einstein-Hilbert action, 
$F=\frac{1}{16\pi G}(R-2\Lambda)$.
If we were to simply ``RG improve'' Einstein's equations,
as in \cite{2007JCAP...08..024B}, we would have
$$
R_{\mu\nu}-\frac{1}{2}Rg_{\mu\nu}=8\pi G T_{\mu\nu}-\L g_\mn
$$
where $G$ is now allowed to depend on $x$.
This is different from the procedure that we followed 
in this paper, which leads to the general equations
\begin{equation}
\label{sonia}
R_{\mu\nu}-\frac{1}{2}Rg_{\mu\nu}=8\pi G T_{\mu\nu}
+G\left(\nabla_\mu\nabla_\nu \frac{1}{G}-g_{\mu\nu}\nabla^2\frac{1}{G}\right)-\L g_\mn
\end{equation}
For want of a better terminology we will refer to these two procedures as 
the ``restricted'' or the ``extensive'' improvement respectively. 
At least at the level of the Einstein-Hilbert truncation it is
not a priori clear that one of these procedures is right
and the other wrong.
For example, one may derive Einstein's equations from arguments
unrelated to an action, then the first choice seems a legitimate one.

Equation (\ref{sonia}) has been written before in
\cite{rw1}, where the restricted improvement is referred to as ``improving equations''
and the extensive improvement as ``improving actions''.  
This terminology is unambiguous in the case of the Einstein-Hilbert action,
but it is ambiguous when one considers more general actions such as (\ref{action}):
in fact is quite clear that also the extensive improvement could very well be described as
``improving equations'', when the equation is written in the form (\ref{motiongeneral}).
Also, while our extensive improvement may well be seen as replacing the constant
couplings by position-dependent couplings in the action,
the procedure that we have followed has been quite different 
from the one advocated in \cite{rw1}, because we have not treated
the position-dependent couplings as prescribed external functions.
Having hopefully clarified the differences in these various approaches,
in the absence of a very strong a priori argument against
the restricted improvement, in this section we will describe the results that
one obtains from it. The equations of motion are of the form (\ref{motiongeneral}) where, now,
\begin{eqnarray}
\label{lhs3}
E_{\mu\nu}
&=&
F'(R)R_{\mn}-\frac{1}{2}F(R)g_{\mn}
-F''(R)\,(\nabla_\mu\nabla_\nu R-g_{\mn}\nabla^2 R)
\nonumber\\
&&
-F'''(R)\,(\nabla_\mu R\nabla_\nu R-g_{\mn}(\nabla R)^2)
\end{eqnarray}

Let us now discuss solutions.
First we observe that in the search of de Sitter solutions,
the difference between the extensive and the restricted improvement is immaterial.
To see this compare the difference between the definition
of $\B$ in the two cases. This difference is given by
\begin{equation}
3\partial_0^2 f'+9H\partial_0 f'
-6\frac{\dot G}{G}\dot f'
-3f'\frac{G\ddot G-2\dot G^2+3HG\dot G}{G^2}
\end{equation}
where $\partial_0$ means that one only takes the derivative
with respect to $t$ of $R$ and not of the couplings.
Each term in this expression contains at least one
time derivative of $H$ and therefore vanishes for de Sitter space. 
Thus, also with the restricted improvement the de Sitter solutions
are given by table 2.

When we look for power law solutions, additional solutions
appear for $n\geq2$.
We consider first the case $n=2$.
\begin{eqnarray}
\A(H)&=&3H^2+18 f_2(2H\ddot H-\dot H^2+6H^2\dot H) 
\nonumber\\
\B(H)&=&6(\dot H+2H^2)+36f_2(H^{(3)}+7H\ddot H+12H\dot H+4\dot H^2)
\end{eqnarray}

One finds that there exist power law solutions with exponents
\begin{eqnarray}\label{pres}
p'_1&=&\frac{4\left(54(1+w)\fast+\xi^2\right)+\frac{2^{4/3}\M'_1}{(\M'_2)^{1/3}}+2^{2/3}(\M'_2)^{1/3}}{6(1+w)\xi^2\left(3-\Last\xi^2\right)}\\ \nonumber
p'_{2,3}&=&\frac{8\left(54(1+w)\fast+\xi^2\right)-\frac{2^{4/3}(1\pm i\,\sqrt{3})\M_1}{\M_2^{1/3}}-2^{2/3}(1\mp i\,\sqrt{3})\M_2^{1/3}}{12(1+w)\xi^2\left(3-\Last\xi^2\right)}
\end{eqnarray}
where
\begin{equation}
\M'_1=4\left(54(1+w)\fast+\xi^2\right)^2-54(1+w)(11+3w)\fast\xi^2\left(3-\Last\xi^2\right)
\end{equation}
and
\begin{eqnarray}\nonumber
 \M'_2&=&16\left(54(1+w)\fast+\xi^2\right)^3-324(1+w)(11+3w)\fast\xi^2\\ \nonumber
 &&\times\left(54(1+w)\fast+\xi^2\right)\left(3-\Last\xi^2\right)\nonumber\\
&&+1944(1+w)^2\fast\xi^4\left(3-\Last\xi^2\right)^2\nonumber\\
 &&+\left[-4(\M'_1)^3+16\left(4\left(54(1+w)\fast+\xi^2\right)^3\right.\right.\\
&&-81(1+w)(11+3w)\fast\xi^2\left(54(1+w)\fast+\xi^2\right)\left(3-\Last\xi^2\right)\nonumber\\ \nonumber
 &&\left.\left.+486(1+w)^2\fast\xi^4\left(3-\Last\xi^2\right)^2\right)^2\right]^{1/2}
\end{eqnarray}

The remarks in the end of section V.B apply here too.
Solutions (\ref{pres}) are all real for sufficiently large $\xi$;
$p_1$ is always negative, while $p_2$ 
diverges for finite $\xi$ and $p_3$ stays almost constant
and is too small for inflation.

For higher truncations the solutions have again to be found numerically.
Figures 6 gives the exponents as functions of $\xi$ in the cases $n=8$ and $n=10$
respectively.
We note that there is a solution that starts at $p\simeq4$ for $\xi=0$
and has the same de Sitter asymptote as the one found with the extensive improvement.
These solutions are very close in the domain of reliability of the truncation.
There is then another solution which starts at $p\simeq 1/2$ 
for $\xi=0$ and has $p>1$ for $3.15\lesssim \xi \lesssim 4.75$,
and whose physical meaning is doubtful.
We conclude that the physically most relevant part of the
solution is rather insensitive to the choice between
extensive and restricted improvement.

%

\begin{figure}
\includegraphics[width=10cm,height=6cm]{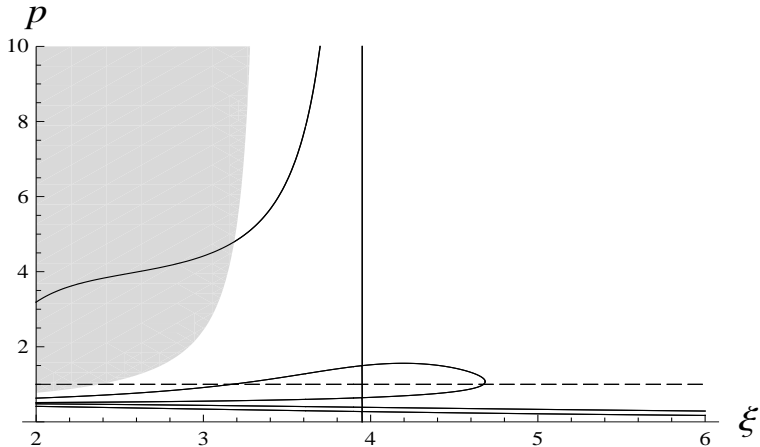}
\caption{Solid curves: numerical solutions for the exponent $p$ 
of power law solutions as a function of $\xi$ for $n=10$, $w=1/3$, 
using the restricted improvement. 
The dashed curve indicates $p=1$. The grey region represents the region excluded by the reliability criterion,
see (\ref{bound}) with $c=1.06$.
}
\end{figure}

\section{Energy and entropy}
One somewhat unsettling aspect of this approach to cosmology is non-conservation
of the energy momentum.
From equation (\ref{motiongeneral}) one finds that
\begin{equation}
\label{conservation}
2\,\nabla_\mu E^\mu{}_\nu=\nabla_\mu T^\mu{}_\nu
\end{equation}
and both sides of the equation would vanish if they were obtained
by varying a diffeomorphism invariant action.
But the RG improved equations were not simply obtained by varying
a diffeomorphism invariant action: after the variation,
the couplings, which are usually treated as constants,
were replaced by functions of the metric.
Of course if we replaced the couplings by scalar functions of the metric
in the action, before varying, then we would obtain another diffeomorphism
invariant action. But this is not the procedure that we use here, so
one should not expect the l.h.s. of (\ref{conservation}) to be zero.
As a consequence, also the r.h.s. cannot be zero, so the energy
momentum tensor on the r.h.s. of the RG improved equations cannot be obtained
from varying some diffeomorphism invariant matter action.

Let us calculate the l.h.s. of (\ref{conservation}).
Since $E_{\mu\nu}$ is linear in $F$, it is easiest to do this
when the function $F$ has a Taylor expansion as in (\ref{seriesF}).
Then one finds
\begin{equation}
\label{bianchi}
\nabla_\mu E^\mu{}_\nu=-\frac{1}{2}\hat\nabla_\nu F
\end{equation}
where $\hat\nabla$ means that the derivative acts only on the couplings
and not on $R$:
\begin{equation}
\label{bianchi2}
\hat\nabla_\nu F=\sum_{i=0}^\infty \nabla_\nu g_i R^i
=\frac{\nabla_\nu k}{k}\sum_{i=0}^\infty \beta_i R^i
\end{equation}
This equation has a very simple interpretation: the failure
of energy-momentum conservation (a gravitational anomaly)
is proportional to the beta functions of the couplings.

At this point it is important to stress that we are not claiming to have
found a violation of energy momentum conservation at a
fundamental level. 
If, as we said in the introduction, we were using the 
full effective action (namely the functional $\Gamma_k$ at $k=0$),
then there would be no RG improvement and,
barring the occurrence of genuine gravitational anomalies
(as discussed for example in \cite{agw}),
there would be no anomaly because the full effective action
is diffeomorphism invariant.
So the type of gravitational anomaly that we are discussing here
is entirely due to the RG improvement and the associated
coarse graining.

To further discuss  this point we can recast (\ref{motionext}) in the form
\begin{equation}
\label{bella}
R_{\mu\nu}-\frac{1}{2}g_{\mu\nu}R= T^{\rm RG}_{\mu\nu} + {\widetilde T}_{\mu\nu} 
\end{equation}
where 
\begin{eqnarray}
\label{incubo}
T^{\rm RG}_{\mu\nu}&=&\frac{1}{f'(R)}
\Big [ \frac{1}{2}g_{\mu\nu}(f(R)-R f'(R)) 
+ \nabla_\mu \nabla_\nu f'(R)-g_{\mu\nu}\nabla^2 f'(R)
\nonumber\\[2mm]
&&
-\frac{\nabla_\mu G}{G}\frac{\nabla_\nu f'(R)}{f'(R)}
-\frac{\nabla_\nu G}{G}\frac{\nabla_\mu f'(R)}{f'(R)}
+2\frac{\nabla^\rho G}{G}\frac{\nabla_\rho f'(R)}{f'(R)}g_{\mu\nu}
\\ \nonumber
&&+G\Big(\nabla_\mu\nabla_\nu\frac{1}{G}-g_{\mu\nu}\nabla^2 \frac{1}{G} \Big)
-\frac{\Lambda}{f'(R)}g_{\mu\nu} 
\end{eqnarray}
and 
${\widetilde T}_{\mu\nu} = ({8\pi G}/{f'(R)}) T_{\mu\nu}$.
It is thus clear that a non-vanishing $T^{\rm RG}_{\mu\nu}$
exactly compensates the nonconservation of $\tilde T_{\mu\nu}$, 
so that the {\it total} effective stress-energy tensor 
$T^{\rm tot}_{\mu\nu}=T^{\rm RG}_{\mu\nu} + {\widetilde T}_{\mu\nu}$ 
given from the r.h.s of (\ref{bella}) is conserved, in general. 

One can think of $T_{\mu\nu}$ as the 
stress-energy tensor for a {\it dissipative}, 
i.e. non-ideal fluid which interacts with the coarse grained
gravitational degrees of freedom.
In fact  in the functional $\Gamma_k$
modes with wavelengths smaller that $2\pi/k$ are integrated over.
Since $k$ is identified with $H$, it decreases with time and therefore,
as time proceeds, more and more gravitational modes are being removed from the 
description of the system.
Such modes carry energy and it should not be surprising
that when they are removed energy seems not to be conserved. 

Although in this paper we have considered the coarse-graining
of the modes of the gravitational field,
similar considerations would apply also to matter fields.
It is interesting to review how this happens already in the 
more familiar framework of Wilson's action for
the $\lambda \varphi^4/4!$ scalar field theory.
Quite generally, the Wilsonian action in this case can be defined as
\begin{equation}\label{block}
e^{-S_k [\Phi]} = \int D[\varphi] \prod_{x} \delta 
(\phi_k(x)-\Phi(x)) e^{-S[\varphi]}
\equiv \int D[\varphi] e^{-S_k[\varphi, \Phi]}
\end{equation} 
where $\phi_k(x)$ is the average of the field $\varphi$ in a domain 
of characteristic length $1/k$. For actual calculation it is possible to introduce
a smearing function $\nu_k(x,x')$ which  is nearly constant within distances
shorter than $1/k$ but rapidly decays to zero outside this region, so that we write 
\begin{equation}
\phi_k(x)= \int d^dx\;\nu_k(x,x') \varphi(x')
\end{equation}
One can evaluate the on-shell condition for the blocked action (\ref{block}) 
by means of the standard saddle-point approximation 
$0=\frac{\delta S_k[\varphi, \Phi]}{\delta \phi}$ 
(see \cite{abe} for details) which gives
\begin{equation}
\Box \varphi = \frac{\lambda}{3!} \varphi^3 +2 M^2 
\Big(\int d^d x' \; \nu_k(x,x') \varphi(x') - \Phi \Big)
\end{equation}
The non-local contribution coming from the coarse-graining kernel
acts as a source terms in the equation of motion for the effective theory at the scale $k$.
This is responsible for a modification of the standard conservation law
which now reads
\begin{equation}\label{rijk}
{\nabla_\nu {T_{\mu}}^{\nu}} = 2M^2 \partial_\mu \varphi (\phi_k - \Phi)
\end{equation}
where $T_{\mu\nu}$ is the stress-energy tensor of the original (bare) field $\varphi$.  
It should be noticed that in the limit $k\rightarrow 0$,  $\phi_k=\Phi$ and one
recovers the standard conservation law with no additional source term.

Let us now see in detail how the modified conservation law arise in the case 
of the Einstein-Hilbert truncation.
The RG improved Friedmann equations read
\begin{eqnarray}
3H^2-3H\frac{\dot G}{G}
&=&8\pi G\r+\L
\\
6(\dot H+2H^2)
-3\frac{G\ddot G-2\dot G^2+3HG\dot G}{G^2}
&=&8\pi G\r\,(1-3w)+4\L
\end{eqnarray}
Taking the derivative of the first equation and using
again both equations, one is led to the following
modified continuity equation
\begin{equation}
\dot\rho+3H(\rho+p)={\cal P}
\end{equation}
where
\begin{equation}
\label{calp}
{\cal P}=
-\frac{1}{8\pi G} \left[\dot\Lambda+8\pi\rho\dot G\right]
-\frac{3 \dot G}{8\pi G^2} \left[\dot H +\frac{H \dot G }{G} + H^2\right]
\end{equation}
The l.h.s. is just $\nabla_\mu T^\mu{}_0$, so this equation 
agrees exactly with the time component of (\ref{conservation}),
when we use the definitions
$$
g_0=-\frac{2\Lambda}{16\pi G}\ ;\qquad
g_1=\frac{1}{16\pi G}\ .
$$
Is is important to stress that this expression can consistently be obtained
from the 4-divergence of Eq.(\ref{bella}), as it must be for consistence. 
Note that we do not use the additional consistency condition described in 
\cite{rw1} as we allow for an unobstructed exchange of energy between matter
and the gravitational effective dynamics. 
The first term in square brackets in the r.h.s of (\ref{calp}) 
had already appeared
in \cite{2007JCAP...08..024B}, but the second one is due to the different
improvement scheme used here. 
We will discuss these differences in some detail the next section.
Here we recall from \cite{2007JCAP...08..024B} that the
time variation of the couplings, and in particular of the
cosmological constant, can be seen as a transfer of energy
to the matter degrees of freedom and hence gives rise to
an increase in entropy of the cosmological fluid.
In fact it was shown that essentially all of the entropy that is observed
can be accounted for in this way.
We note that this calculation, which is given in equations (3.9)-(3.15) in \cite{2007JCAP...08..024B},
does not depend on the form of ${\cal P}$ as a function of
$\dot \Lambda$ and $\dot G$, so the entropy generation will
be the same in the approach followed here. In fact, as discussed in \cite{2007JCAP...08..024B},
the peak of the entropy production is in the crossover region, near the Gaussian fixed point, where
only the running of the cosmological constant is relevant and $\dot G \sim  0$.  

We now give the formula for energy (non)conservation in the general case.
In general we can write
$$
\nabla_\mu E^\mu{}_0=-\partial_\mu E_{00}-4H E_{00}-HE^\mu_\mu\ .
$$
Then using
$$
E_{00}=\frac{\A(H)-\Lambda}{16\pi G}\ ;\qquad
E^\mu_\mu=-\frac{\B(H)-4\Lambda}{16\pi G}
$$
and the explicit expressions (\ref{Adef},\ref{Bdef}), one finds
$$
{\cal P}=2\nabla_\mu E^\mu{}_0
=-\frac{1}{8\pi G}
\left[
\dot\Lambda-\Lambda\frac{\dot G}{G}
-\frac{1}{2}(\dot f-f'\dot R)
+\frac{1}{2}f\frac{\dot G}{G}\right]
$$
This agrees with the result of inserting (\ref{parametrization})
into (\ref{bianchi}).

We can see the implications of these facts
for the search of cosmological solutions at the fixed point.
At a fixed point we have
$$
\beta_i=(4-2i)g_i
$$
so inserting in (\ref{bianchi}) we find
$$
\nabla_\mu T^\mu{}_\nu
=2\nabla_\mu E^\mu{}_\nu
=-\frac{\nabla_\nu k}{k}\sum_{i=0}^\infty  (4-2i)g_i R^i
=-2\frac{\nabla_\nu k}{k}(2F-R F')
$$
There are therefore two ways in which the anomaly could vanish:
the first is that $2F=R F'$, which is satisfied if and only if $F=g_2 R^2$.
This is because $g_2$ is dimensionless, and therefore it is
constant at a fixed point.
The other way, which could work for any form of the function $F$,
is that $\nabla_\nu k=0$.
This depends on the choice of the cutoff, i.e. how it depends on the metric,
and on the solution.
In particular, if we look for vacuum solutions $T_{\mu\nu}=0$,
we must also have $\nabla_\mu T^\mu{}_\nu=0$.
Vacuum solutions must have a vanishing anomaly and this is why
they are harder to come by than solutions with matter
\footnote{The same argument holds more generally if we demand that the
energy momentum tensor be derivable from a diffeomorphism invariant matter action.}.
If we make the cutoff identification $k=\xi H$,
and if $F$ is not just $g_2 R^2$, then a necessary condition to have a
vacuum solution is $\dot H=0$, so the only vacuum solution is de Sitter.

The (non)conservation of the energy momentum tensor with a restricted improvement works in a different way.
Instead of (\ref{bianchi}) we find now
\begin{eqnarray}
\nabla_\mu E^\mu{}_\nu&=&R_{\mu\nu}\hat\nabla^\mu F'
-\frac{1}{2}\hat\nabla_\nu F
-(\nabla_\mu\nabla_\nu R-g_{\mu\nu}\nabla^2R)\hat\nabla^\mu F''
\nonumber\\
&&
-(\nabla_\mu R\nabla_\nu R-g_{\mn}(\nabla R)^2)\hat\nabla^\mu F'''(R)\,
\end{eqnarray}
Note that only the second term is present if we perform the extensive improvement.
The new terms that are seen here can be viewed as additional
contributions to the anomaly.

Finally let us discuss the energy density that is required
for the existence of the power law solutions.
Once the solution for $a(t)$ is found, one can plug it in
(\ref{eqn2rho}) to obtain $\rho$.
It turns out that $\rho$ always depends on time as $t^{-4}$,
as demanded by dimensional analysis, so the only issue is the
value of the constant prefactor.
Explicit formulae are easily derived in the case $n=1$. One finds
\begin{eqnarray}
\rho&=&\frac{\A(H)-\Lambda}{8\pi G}\nonumber\\
&&=-\frac{9\xi^2}{2\pi\Gast}\frac{3w+\Last\xi^2\pm\sqrt{3(1+w)\left(3(w-1)+2\Last\xi^2\right)}}{(1+w)^2\left(3-\Last\xi^2\right)}\frac{1}{t^4}
\end{eqnarray}
in the extended RG improvement scheme and
\begin{equation}
\rho=\frac{3H^2-\Lambda}{8\pi G}
=\frac{81\xi^2}{128\pi\tilde G_*(3-\tilde\Lambda_*\xi^2)^3}\frac{1}{t^4}
\end{equation}
in the restricted improvement scheme.
Note that the additional term proportional to $\dot G$,
which is present in the former case, is negative, 
making the density  $\rho$ negative in the former case and positive in the latter.

These facts signal that a detailed analysis of the matter sector is necessary. 
We shall not undertake this analysis here, leaving it for future work. 
We just add a few remarks.
It is not difficult to see that one can have positive energy
also with the extended improvement, provided $w$ is smaller than a critical 
value $w_{\rm cr}<-1$.
Furthermore, in a more refined treatment we should perform the 
coarse-graining also on the matter, treated as quantum fields. 
On the other hand, 
it can be argued that the realistic RG trajectory 
is the one which emanates from the non-Gaussian FP
and spends a long time near the Gaussian fixed point, 
corresponding to the classical era \cite{2007JCAP...08..024B}. 
During this time  $\dot G \sim 0$ and one recovers the more familiar
``restricted'' RG evolution where the density is always positive.  


\section{Discussion}

The results obtained here can be summarized by saying that if we identify
the cutoff with a multiple of the Hubble parameter, as in (\ref{cutoff}),
inflationary power law solutions (i.e. with exponent $p>1$) exist for $2.7\lesssim\xi\lesssim3.9$.
The dependence on $\xi$ is strong, with the exponent diverging at $\xi\approx3.9$.
Exactly at that point, the theory admits a de Sitter solution.
Provided that $\xi$ lies in the above range, 
that the starting point is close enough to the FP and that $p>1$,
it should always be possible to have a sufficient number of $e$-foldings.

Weinberg \cite{weinberginflation} has studied the FRW equations
that follow from a general gravitational action containing arbitrary
powers of curvature, and the possibility that they admit inflationary
(de Sitter) solutions.
In his approach the cutoff is a fixed mass scale that has
to be optimized for the treatment of inflation.
Unlike the approach followed here, it does not depend on time
and therefore there are no ``RG improvement'' terms in the equations.
This may sound like a very different procedure, but in practice it is not,
for two reasons.
The first is that the fixed optimal cutoff is tuned to a description
of inflation, and if one wanted a cutoff that is tuned to some
later stage of the cosmic history, it would be different;
thus effectively one would have again a time-dependent cutoff.
Conversely, when we focus only on de Sitter solutions,
also our time-dependent cutoff $k=\xi H$ becomes time-independent.
Therefore, if one started with the equations of \cite{weinberginflation}
and truncated the action to the form $F(R)$ that we consider here,
then one would find the same solutions.

The equations of motion determine $H$ as a function of
the cutoff, so if we assume that the cutoff is a multiple of $H$, the
coefficient $\xi$ is determined by the solution, as we have observed.
It is worth noting that the value of $\xi$ that produces a de Sitter
solution is in the right physical range, namely the cutoff retains
fluctuations with wavelengths that are few times smaller that the horizon scale.

In this work we have discussed the effect of performing the RG improvement
in the equations of motion, after having varied the action.
To see how things could work in a different approach, let us momentarily
view equation (\ref{expeq}) as a differential equation for $f$
(instead of an algebraic equation for $R$, as we have looked at it so far).
It has the solution $\Lambda=0$, $f=g_2 R^2$, which corresponds to
a scale invariant theory.
(We have also seen in section VII that in this case the anomaly automatically vanishes.)
For such an $f$, de Sitter space is a solution for any value of $R$,
and $\xi^2=12/\tilde R$ remains undetermined.
We observe that this theory can be obtained if we make the RG improvement
in the action, before varying.
In fact, if we make the cutoff identification $k^2=\omega R$,
in the fixed point regime $g_i=\tilde g_i \omega^{4-2i}R^{2-i}$, so
$$
\sum_{i=0}^\infty g_i R^i=\bar g_2 R^2\ ;\qquad
\bar g_2=\sum_{i=0}^\infty \tilde g_i\omega^{2-i}
$$
Note that since the fixed point coefficients $\tilde g_i$ seem to be
all less than one, and to have alternating signs, with $\omega>1$ there is
reasonable hope that the series converges. 
The same behavior obtains with the cutoff identification (\ref{cutoff}),
if one restricts onself to de Sitter spaces.
This result also agrees with the expectation that the fixed point should be described
by a scale invariant action.
Cosmological solutions of this theory have been studied in \cite{menegoz}.

There are properties of the theory that are not universal,
but depend on the choice of the cutoff.  
Clearly, it will be important to check that the observable low-energy physics is not strongly 
dependent on the regulator choice, but this question can only be addressed with a proper numerical
integration of the complete system of improved RG equations and the $\beta$-functions for 
the couplings. 

One aspect of the problem that we have not touched upon here is the
end of inflation.
In \cite{weinberginflation} this is estimated by considering the
development of instabilities in the exponential solution.
In the approach that we followed here one would assume that the
initial point of the RG evolution is not {\it at} the fixed point
but somewhere close to it.
The solutions described here would then hold only approximately.
The RG evolution would take the theory away from the fixed point,
with a speed that can be calculated from knowledge of the
scaling exponents of the theory.
Eventually when one gets sufficiently far from the fixed point the
inflationary solution would give way to an ordinary radiation
dominated universe.
One would think that by choosing the initial point sufficiently close
to the fixed point one can have an inflationary period of arbitrarily
long duration.
Of course the two approaches should give at least similar results.
We hope to return to this point in the future.

Unfortunately the results of this paper are not yet a realistic basis
for a model of inflation.
The main reason is that we are neglecting a great number of terms in the action.
For the study of Robertson-Walker cosmologies, Weyl terms are unimportant,
and for $n=2$ (meaning with four derivatives) there is nothing beyond $R^2$
that is not a total derivative. However from $n=3$ upwards there are 
many terms in the action that contain traces of powers of the Ricci tensor,
whose effect we are currently unable to estimate.
One rather sobering result of our analysis has been
that in order to find reliable solutions one has to go to really
high orders in the derivative expansion.
It is encouraging, however, that the (technically unreliable) results
of the Einstein-Hilbert truncation \cite{2007JCAP...08..024B} proved in the end to
give the correct qualitative picture (at least within the class of $F(R)$ truncations).
One may hope that also when Ricci terms are included, the results
of the low order truncations are not too misleading.

\end{document}